# An Object-Oriented Metamodel for Bunge-Wand-Weber Ontology

**Arvind W. Kiwelekar, Rushikesh K. Joshi**
Department of Computer Science and Engineering
Indian Institute of Technology Bombay
Mumbai 40076, India
{awk,rkj}@cse.iitb.ac.in

## Abstract

A UML based metamodel for Bunge-Wand-Weber (BWW) ontology is presented. BWW ontology is a generic framework for analysis and conceptualization of real world objects. It includes categories that can be applied to analyze and classify objects found in an information system. In the context of BWW ontology, the metamodel is a representation of the ontological categories and relationships among them. An objective behind developing an object-oriented metamodel has been to model BWW ontology in terms of widely used notions in software development. The main contributions of this paper are a classification for ontological categories, a description template, and representations through UML and typed based models.

## 1 Introduction

Applications of ontology in formalizing semantics of modeling language constructs [Joerg, 2005; Yair and Weber, 1990; Joerg and Wand, 2005; Yair and Weber, 1999; Green *et al.*, 2005], in knowledge representation [Sowa, 2000], and in modeling information systems [Uschold *et al.*, 1998] show the growing interest of software developers and modelers in this branch of philosophy. Ontology is concerned with general features and facts about the real world. In ontology, we seek to answer philosophical questions like- 'what kinds of objects are found in the real world' and 'how these objects are organized'. Few examples of general ontologies are General Formal Ontology [Heller and Herre, 2004] and Bunge's ontology. Scope of this paper is restricted to the ontology of Bunge-Wand-Weber (BWW) [Bunge, 1977; Yair and Weber, 1993; 1995]. Bunge's ontology serves as a foundation for BWW ontology. Bunge's original ontology [Bunge, 1977] is considered as a general system theory. Wand and Weber [Yair and Weber, 1993; 1995] have adapted it to model information systems.

The postulates in BWW ontology are widely accepted statements about real world phenomena and are based on everyday experiences, observations and facts. In [Bunge, 1977], semantics of ontological categories are formalized through set theoretic notations. Later, Wand and Weber [Yair and Weber, 1993; 1995] followed the same approach to formalize ontological categories. However, as noted by Rosemann and Green [Michael and Green, 2002] BWW ontology has received criticism from the point of view of lack of understandability, comparability and applicability. In order to bridge this gap, an Entity-Relationship (ER) based metamodel was presented in [Michael and Green, 2002]. Continuing the same argument, the work reported in this paper attempts an Object-Oriented (OO) metamodel for a small subset of BWW ontology that is relevant to software system modeling.

### 1.1 Approach

The following issues were considered during the development of this metamodel.

1. Can the ontological categories be grouped according to some criteria?
2. How are the ontological categories related to each other?
3. How to formalize a visual model in a formal notation understandable to software practitioners?

As a first step to represent BWW ontology through OO based metamodel, ontological categories are classified into different groups. Secondly, the metamodel is attempted to capture the relationships among ontological categories. The metamodel is represented through two different models i.e visual and descriptive models. The visual model for BWW ontology is represented through UML [OMG, ] notations. Thirdly, the notion of *simple and composite types*, is used for descriptive modeling. In the descriptive modeling, an ontological category is interpreted as a *type*. Supplementary functions in the form of predicates are defined to capture the constraints on relationships. A template has been defined for descriptive modeling and is uniformly applied to describe the ontological categories.

The rest of the paper is organized as follows. Section 2 proposes a classification ontological categories. In Section 3 the guidelines and conventions followed for visual and descriptive modeling are specified. The ontological categories: intrinsic, relational, compositional and collection categories are discussed in subsequent sections.

## 2 Classifying Ontological Categories

Ontological categories capture real world phenomenon and organize objects found in the real world. BWW ontological categories are generic in the sense that they are not restricted to one particular domain as in Enterprise Ontology [Uschold *et al.*, 1998]. Table 1 depicts BWW ontological categories classified in five different groups. This classification is intended to improve our understanding of the nature of these categories and relationships between them.

| Sr. No. | Types of Categories | Examples of Ontological Categories |
|---|---|---|
| 1. | Intrinsic Categories | Property, Thing and State |
| 2. | Representational Categories | Attribute, Schema and State Variable |
| 3. | Primitive Relational Categories | Possesses, Precedes and Event |
| 4. | Composition Categories | Conjunction, Association, Event Composition |
| 5. | Collection Categories | Class, Kind, Natural Kind, State Space and History |
| 6. | Supplementary Categories | Intrinsic Property, Mutual Property, Binding Mutual Property, Non-Binding Mutual Property, Part-of, Emergent Property, Resultant Property, Actson, Internal Event, External Event, Coupled Event, Subclass, Sub Kind, System, Structure, and Environment |

Table 1: **BWW Ontological Categories**

1. **Intrinsic Categories** The ontological categories included in this group are Property, Thing and State. These categories are called *intrinsic* because these are the most significant and fundamental one. Bunge's postulates [Bunge, 1977] that are captured through these categories are *World consists of things possessing properties*, and *Every thing changes*.

2. **Representational Categories** The ontological categories from this group are used to describe a real world phenomena. An intrinsic category and a descriptive category from this group are related through a *representation* relationship. For example, Schema represents Thing, Attribute represents Property, and State variable represents State.

3. **Primitive Relational Categories** These are simple binary relations relating two *intrinsic* ontological categories. For example, *possesses* relates things and properties, *Precedes* relates properties to themselves.

4. **Composition Categories** These are categories defined to construct a complex category from simple categories. For example *conjunction* composes complex properties out of simple properties, *association* composes a complex or composite thing from smaller things, and *event composition* defines a process i.e. a complex event.

5. **Collection Categories** The purpose of ontological categories from this group is to collect related objects together to form a collection. The examples are Class, Kind, State Space and History as an ordered collection.

6. **Supplementary Categories** The ontological categories from this group are dependent on the categories defined earlier. Few examples of these categories are resultant property, emergent property, and actson. However, these categories have not been considered in this paper. They can be further classified in different groups.

## 3 Representational Conventions

This section briefly explains the notational conventions followed for representing the metamodel. Two different representations of the metamodel are attempted i.e. i) The visual model that pictorially depicts relationships among ontological categories. ii) The descriptive model that represents metamodel in a formal way by specifying invariants.

### 3.1 Visual Representation

A rectangular box i.e. UML symbol for a classifier is used to represent an ontological category. The name of category is displayed inside the classifier box. A generalization category is denoted through a thicker classifier box than that of concrete categories. Relationships among categories are represented as UML associations. Figure 1 shows the scheme of representing a binary relationship. Sometimes, a relationship further participates in other relationships. This fact is represented through association classes. Figure 2 shows a scheme for representing relationship as an association class. In all, the metamodel uses UML notations for association, aggregation and generalization.

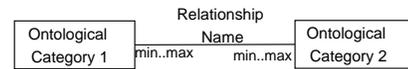

Figure 1: **Scheme for representing a Binary Relationship**

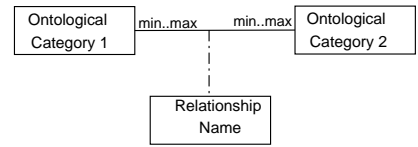

Figure 2: **Relationship as an Association Class**

### 3.2 Representing Ontological Categories as types

The visual model represents relationships among ontological categories and constraints like cardinalities. In addition to this, invariant properties are also defined. An invariant is a defining characteristic for a relationship. To define an invariant, *supplementary functions* defined over the ontological category type are used. Types used are *basic types*, *product types* and *mappings*. Constraints on relationships can be specified in terms of supplementary functions. Two standard types i.e. *Boolean* and *Time* are used in the description.

| Sr. No. | Thing | Properties | States | Schema |
|---|---|---|---|---|
| 1. | Book as a Library Item | Title, Author, Price, ISBN, Publisher, Classification No. | onTheRack, issued, claimed, written-off, missing, | *Book*(Class. No, Title, Author), *Book*(ISBN, Title, Publisher) |
| 2. | University Student | Reg. No. Names, Address, Date of Birth, Course Registered, Degree Awarded, | registered, graduated, migrated | *Student*(Reg.No, Name, Address) *Student*(Reg. No., Course Registered, Degree awarded) |
| 3. | Cricket Player | Name, Runs Scored, Centuries Scored, Wickets Taken, 5-wicket Hauls, | playing, injured, rested, retired | *Player*(Name, Runs Scored, Centuries scored) *Player*(Name, Wickets Taken, 5-wicket haul) |
| 4. | Network Printer | Name, Make, Location | on, off, busy, idle | *Printer*(Name, Make, Location) |

Table 2: **Examples of Intrinsic and Representational Categories**

### 3.3 Descriptive Template

Each relationship is characterized through a template that involves the following elements.

- **Intention** This element describes an observation or a fact that a category tries to capture.

- **Participating Categories** The categories participating in a relationship are specified through type signature.

- **Examples and Non-Examples** An example illustrating the phenomenon intended to be captured through the concerned category is given. Also, to further clarify meaning of an ontological category a close non-example is provided.

- **Supplementary Categories** The names of ontological categories that are derivable from the concerned category are given.

- **Invariant** This item is applicable for supplementary and relational categories. An invariant characteristic is defined for the concerned category in terms of constraints on relations with other categories.

The following symbols are used.

| | | | |
|---|---|---|---|
| $X$ | Relational Type (cross product) | $\rightarrow$ | Function Type (mapping) |
| :: | definition | = | Equality test operator |
| $\Leftrightarrow$ | equivalence (bidirectional implication) | | |

*Supplementary Functions:* Each supplementary function is suffixed either by a $'?'$ or a $'!'$ symbol. Symbol $'?'$ denotes that the supplementary function is intended to test satisfiability of a particular condition. The symbol $'!'$ denotes that the supplementary function is a correspondence function.

### 3.4 Scope of the Paper

The metamodel discussed in this paper is intended to represent the intrinsic, representational, primitive relational, compositional and collection categories. *System* related ontological categories, which have been identified as supplementary categories are not discussed in this paper.

## 4 Intrinsic Categories

Intrinsic categories like property, thing and state are the central notions in BWW ontology capturing both static and dynamic features of objects found in reality. The following table shows the type description and a supplementary function for this class of categories.

| Intrinsic Types | Property, Thing, State |
|---|---|
| Special Element | null :: Thing |
| Supplementary Function | $isIn? :: Thing \ X \ State \ X \ Time \rightarrow Boolean$ |

### 4.1 Property

The notion of a property characterizes objects found in reality. Properties capture static and dynamic features of an object.

- **Intention** To capture the fact that *Objects have properties*

- **Examples** Table 2 gives examples for the category Property.

- **Non-Examples** Things to which properties are associated are not the examples of properties. In BWW ontology, properties as individuals do not have any existence. *Whiteness* as a property does not have any existence. A paper is a thing possessing whiteness property.

- **Supplementary Categories** Intrinsic Property, Mutual property, Emergent, Resultant and Complex Property.

The following section discusses intrinsic and mutual properties.

**Intrinsic and Mutual Property**

A *dependence* relationship is used to distinguish between two types of properties i.e. *intrinsic property* and *mutual property*. (i) An intrinsic property is a property that is dependent on a single object. For example, age and height of a person are the intrinsic properties. (ii) Mutual properties are also known as relational properties. For example, $worksFor$ is a relational property between *employee* and a *company*. Mutual properties are further classified as binding and non-binding properties. An *interaction* relationship between two things is considered to classify a mutual property. (a) In case of a non-binding mutual property, no interaction is involved between two related things. For example, *younger than* relationship between two persons does not show any kind of interaction. (b) On the other hand, two persons may relate with each other through a *sales* relationship i.e. one person buying a product from another is an example of a binding property.

- **Non-Examples** A simple property or an attribute does not represent a state. The value assigned to a particular attribute is a state, for example attribute i.e. $Availability\_status = onTheRack$ implies that certain book is on the rack and is available for issue. Attribute $Availability\_status$ is not a state but $onTheRack$ is a state.

The supplementary function $isIn?(x, s, t)$ verifies whether thing $x$ is in a state $s$ at time $t$.

## 5 Representational Categories

The representational categories like Schema, Attribute and State Variable are used to describe an intrinsic category. As shown in Figure 4, thing is represented by schema, property is represented by attribute, and state is represented by state-variable. A compact description of the representational categories is given below.

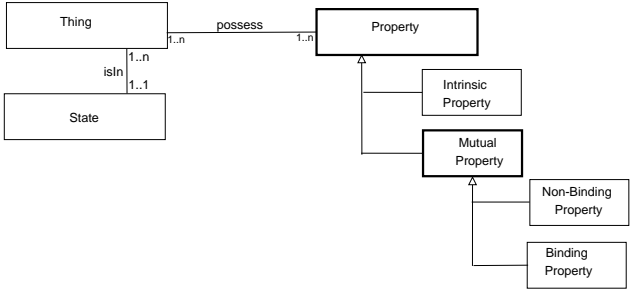

Figure 3: **Intrinsic Categories**

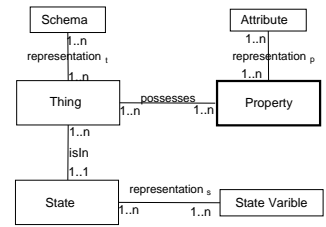

Figure 4: **Representational Categories**

### 4.2 Thing

Thing is a substantial object having existence in reality.

- **Intention** To capture the fact that *World consists of things.*
- **Examples** Table 2 gives examples for the category Thing.
- **Non-Examples** A fictitious object like *Superman* in the real world is not considered as a thing. Since things are not mere bundles of properties but should have a physical existence that possesses these properties. However, *Superman* as a character in a movie is a valid thing because it has physical existence in its domain.
- **Supplementary Categories** Composite thing and System.

A special thing called $null$ is pre-defined in BWW ontology.

### 4.3 State

The notion of a state is based on the postulate that every thing is in some state or the other at a given time.

- **Intention** To capture changing nature of a thing.
- **Examples** Table 2 gives examples for the category State.

- **Intention** To describe an intrinsic category in multiple ways.
- **Examples** (i) Table 2 gives examples for the category Schema. (ii) A property like $Address$ is represented by attributes *House Number*, *Street* and *City*. (iii) State is represented either by a single state variable or a bunch of state variables. State variable is a function that maps a property of a thing to a specific value in certain codomain, such as in $statusOfBook :: Book \rightarrow Boolean$.
- **Non-Examples** (i) Schema is not an actual description of an object with values like $(123, "goodbook", "anauthor")$. Schema is a thing-specific general descriptive framework. (ii) The properties like *fingerprint, blood group* possessed by a person or ISBN of a book are not the examples of state variables.

## 6 Primitive Relational Categories

This section describes primitive relational categories. In the following table, these categories are represented as types and supplementary functions are defined.

| **Relational Types** |
| --- |
| $Possesses :: Thing \; X \; Property$ |
| $Precedes :: Property \; X \; Property$ |
| $Event :: State \; X \; State$ |
| **Supplementary Functions** |
| $possesses? :: Thing \; X \; Property \rightarrow Boolean$ |
| $precedes? :: Property \; X \; Property \rightarrow Boolean$ |
| $event? :: Thing \; X \; State \; X \; State \rightarrow Boolean$ |
| $fromState! :: Event \rightarrow State$ |
| $toState! :: Event \rightarrow State$ |
| $composibleEvent? :: Event \; X \; Event \rightarrow Boolean$ |

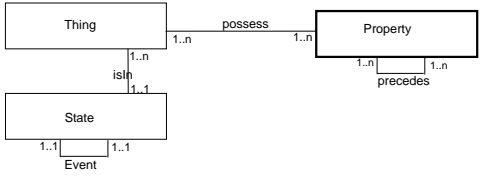

Figure 5: **Primitive Relationships**

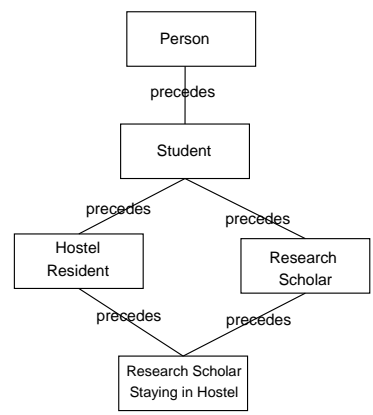

Figure 6: Example for Precedence Relationship

## 6.1 Possesses

This is a relationship between things and properties. A supplementary function, $possesses?$, is defined to test whether thing $t$ possesses property $p$.

- **Intention** To capture the fact that *all things possess properties*.

- **Participating Categories**

    $Possesses :: Thing \; X \; Property$

- **Examples** *Book* as a thing possesses properties like *title*, *author*, *publisher* etc.

- **Non-Examples** Hard-disk is not a property that is possessed by computer.

In *possesses* relationship, minimum cardinality assigned to Thing and Property is 1, indicating that there is no such thing like- a thing without a property and a property without a thing.

## 6.2 Precedes

This is a relationship among properties. A supplementary function $precedes?(p_1, p_2)$ is used to verify whether property $p_1$ precedes $p_2$.

- **Intention** To capture the fact that one property is a necessary condition for another one.

- **Participating Categories**

    $Precedes :: Property \; X \; Property$

- **Examples** Figure 6 shows an example of properties that can be constructed through precedence relationship. In this example, "being a person", "being a student", "staying in a hostel" etc. are properties related through precedence pairs.

- **Non-Examples** Properties "age as 10" and "being a vegetarian" are not related through precedes relation.

The *precedes* relationship is a *reflexive* and *transitive* relationship.

## 6.3 Event

This is a relationship between two states of a thing. A supplementary function, $event?(z, s_1, s_2)$ is defined to test that there exists a change in a thing $z$ from state $s_1$ to $s_2$.

- **Intention** To capture the fact that *all things change*.

- **Participating Categories**

    $Event :: State \; X \; State$

- **Invariant** A thing is said to have undergone a change if the thing is in two different states at two different instances of time, and there is no other state between the two.

$$event?(z, s_1, s_2) :: Thing(z) \wedge State(s_1) \wedge State(s_2) \wedge$$

$$\exists t_1, t_2 (Time(t_1) \wedge Time(t_2) \wedge isIn?(z, s_1, t_1) \wedge$$

$$isIn?(z, s_2, t_2) \wedge (t_1 < t_2) \wedge s_1 \neq s_2$$

$$\wedge \neg \exists s, t_3 (State(s) \wedge Time(t_3) \wedge (t_1 < t_3 < t_2) \wedge isIn?(z, s, t_3)$$

$$s \neq s_1 \wedge s \neq s_2))$$

- **Examples** When a Library Book is borrowed by some student, its state changes from *onTheRack* to *Issued*.

- **Supplementary Categories** Process, Actson, Coupled Event and Transformation .

A supplementary function that tests whether two events are compatible or not is defined below.

$composibleEvent?(e_1, e_2) :: toState!(e_1) = fromState!(e_2)$

# 7 Composition Categories

In BWW ontology, three composition categories are defined to form a complex object out of simple objects. These are *conjunction*, *association* and *event composition*. In the following Table, these categories are represented as types and the supplementary functions are defined.

| Composition Types |
|---|
| Complex Property :: $Property^n$, n>0 (n-conjuncts) |
| Association :: $Thing^n$, $n > 0$ (n-ary association) |
| Event Composition :: $Event^n$, n>0 (n-step process) |
| **Supplementary Relational Types** |
| $Partof :: Thing\ X\ Thing$ |
| **Supplementary Functions** |
| $composite? :: Thing \rightarrow Boolean$ |
| $complexProperty? :: Property \rightarrow Boolean$ |
| $process?() :: Event \rightarrow Boolean$ |
| $partof? :: Thing\ X\ Thing \rightarrow Boolean$ |

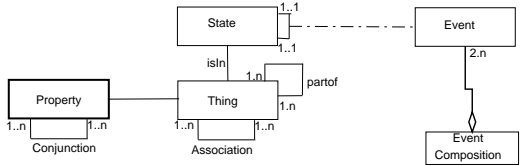

Figure 7: Composition Categories

## 7.1 Complex Property

More than one property is combined to form a *complex property* through conjunction. Conjunction defines composition of properties. In the metamodel, conjunction is represented as n-ary relationship between properties.

- **Intention** To capture the assumption that *properties combine with each other to form a complex property*.

- **Participating Categories**

$$Conjunction :: Property^n$$

$Property^n = Property_1\ X\ Property_2\ ..\ X\ Property_n$. The resultant category is called *Complex property*. Complex property is the conjunction of $1\ to\ n$ properties.

- **Examples** In Figure 6, the property "Research Scholars staying in Hostel" is a complex property combining "Hostel Resident" and " Research Scholar".

- **Non-Example** Intrinsic properties like finger-print and blood-group are not the examples of complex properties.

A supplementary function that tests whether a given property is a complex property is defined below.

$$complexProperty?(p) :: \exists p_1, p_2(Property(p_1) \wedge$$
$$Property(p_2) \wedge (p = p_1 \wedge p_2))$$

## 7.2 Association

Association in BWW ontology is a compositional relation. It is intended to compose simple things to form one *composite thing*. In the metamodel, association is represented as n-ary relationship between things. An association of things is a new thing with an identity.

- **Intention** To capture the assumption that *things associate with each other to form a composite thing*.

- **Participating Categories**

$$Association :: Thing^n$$

where $Thing^n = Thing_1\ X\ Thing_2\ ..\ X\ Thing_n$. The resultant category is called *Composite Thing*. A composite thing is an association of $1\ to\ n$ things.

- **Examples** Network of workstations is an association of workstations.

- **Non-Examples** The relationships like $brotherof$, $worksfor$ are not the examples of association. These are relational or mutual properties.

- **Supplementary Categories** Part-of.

In the following subsection *Part-of* relation is discussed.

**Part-of**

The part-of relationship is a supplementary relationship in Bunge's ontology. A supplementary function, $partof?(x, y)$, tests whether $y$ is a part of $x$.

- **Intention** To capture the fact that a large thing is composed of several small things.

- **Participating Categories**

$$Partof :: Thing\ X\ Thing$$

- **Examples** A hard disk is part-of a personal computer.

- **Non-Examples** A property is not a part-of a thing. For instance, when a person drives a vehicle, $drivenBy$ is a mutual property and neither the person nor the vehicle are parts of each other.

## 7.3 Event Composition (Process)

Event composition in BWW ontology composes events to form a *complex event*. In the metamodel, event composition is represented as n-ary relationship between events.

- **Intention** To capture a complex change in terms a sequence of smaller events.

- **Participating Categories**

$$Event\ Composition ::\ Event^n$$

where $Event^n = Event_1\ X\ Event_2\ ..\ X\ Event_n$. The resultant category is called *Process*. Process is the composition of $1\ to\ n$ events.

- **Examples** In the case of a Book as a library item, a pair of events <issued,claimed> and <claimed,issued> forms a process.

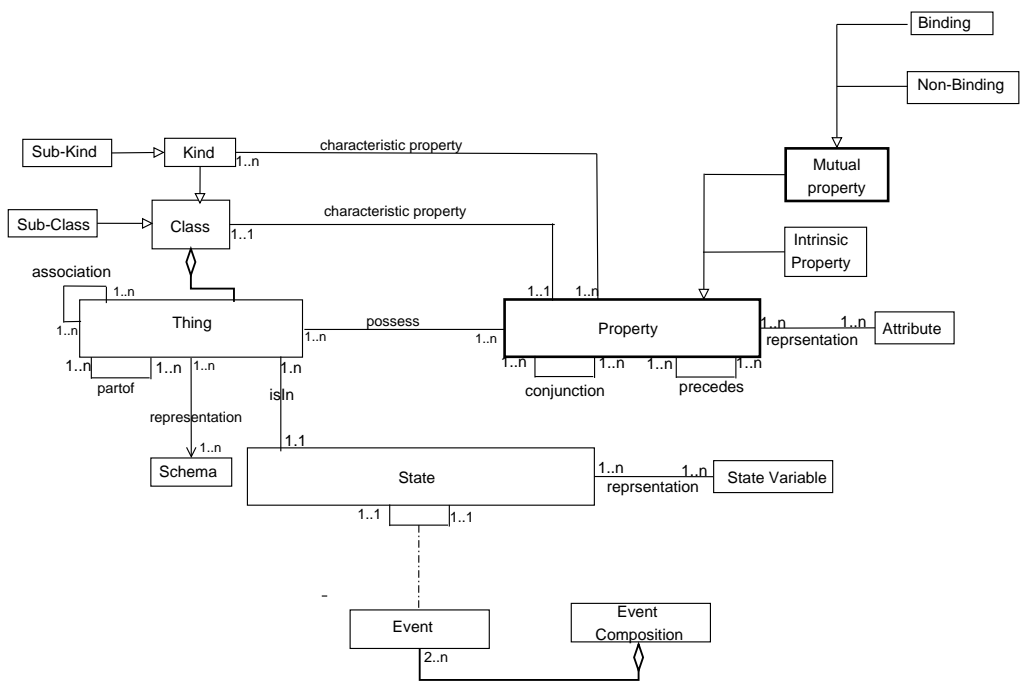

Figure 8: **Metamodel for BWW Ontology**

## 8 Collection Categories

The collection categories allow us to group related objects together and treat the group as a category. In the following table, collection categories are represented as types and the supplementary functions are defined. Figure 9 shows the relationships among collection categories.

| Collection Types |
|---|
| $Class = 2^{Thing}$ |
| $Kind = 2^{Thing}$ |
| **Supplementary Functions** |
| $memberof_c? :: Class \times Thing \rightarrow Boolean$ |
| $class? :: 2^{Thing} \times Property \rightarrow Boolean$ |
| $kind? :: 2^{Thing} \times 2^{Property} \rightarrow Boolean$ |
| $characteristicProp_c? :: Class \times Property \rightarrow Boolean$ |
| $characteristicProp_k? :: Class \times 2^{Property} \rightarrow Boolean$ |

### 8.1 Class

Class category groups similar *things* together. The fact that class is not any arbitrary collection of things is captured through a characteristic property. Since an example class is a set of its instances, the category Class is a power-type. A supplementary function, $memberof_c?(C,t)$, to test whether thing $t$ is member of a class, a collection $C$, is defined.

- **Intention** To group all things that possess a certain property. A property that is possessed by all things in a class is called *characteristic property*.

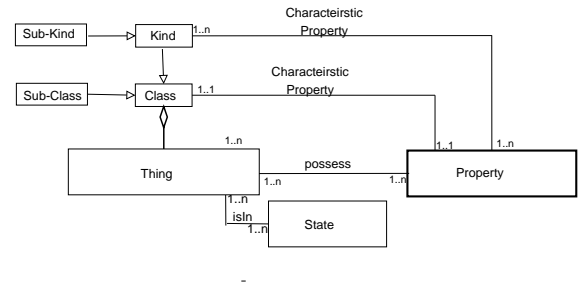

Figure 9: Collection Categories

- **Participating Categories**

$$Class :: 2^{Thing}$$

- **Invariant** All things which are members of a collection that is class possess the characteristic property of the class. Also, there is no thing outside the class possessing the same property.

$$class?(c,p) :: c \in 2^{Thing} \land Property(p) \land$$

$$characteristicProp_c(c,p)? \land$$

$$memberof_c?(c,t) \Leftrightarrow possesses?(t,p)$$

To every class exactly one characteristic property is assigned.

- **Example** A university student possesses a property called *Enrolled*.
- **Non-Example** Any arbitrary collection of things like horse, television, paper, etc is not a class.
- **Supplementary Categories** Subclass

## 8.2 Kind

$Kind$ category organizes things through a set of properties.

- **Intention** To further organize things in an orderly fashion based on a set of properties.
- **Participating Categories**

$$Kind :: 2^{Thing}$$

- **Example** Child labor is a set of persons having properties $underaged$ and $works for(p, c)$. The first one is an intrinsic property, while the second one is a mutual property.
- **Supplementary Categories** Subkind

A kind may contain a few instances from many classes since it uses more than one property to define its set of instances.

## 9 Conclusions and Future Work

An object-oriented metamodel depicting ontological categories and relationships among them was attempted. The same is summarized in Figure 8. The main highlights of the metamodel presented in this paper are: (i) Classification of BWW ontological categories to improve understanding of categorization. (ii) Representing metamodel through visual model to count on familiarity, and (iii) Capturing the constraints by modeling categories as types in anticipation of closeness to implementation. Further modeling for supplementary categories and system related categories is being carried out.

## References

[Bunge, 1977] Mario Bunge. *Treatise on Basic Philosophy (Vol 3): Ontology I : The Furniture of the World*. D. Reidel Publishing Compant, first edition, 1977.

[Green *et al.*, 2005] Pete F. Green, Michael Rosemann, and Marta Indulska. Ontological evaluation of enterprise systems interoperability using ebxml. *IEEE Transactions on Knoweledge and data Engineering*, 17(5):713–724, May 2005.

[Heller and Herre, 2004] Barbara Heller and Heinrich Herre. Ontological categories in gol. *Axiomathes*, 14(3):57–76, 2004.

[Joerg and Wand, 2005] Evermann Joerg and Yair Wand. Toward formalizing domain modeling semantics in language syntax. *IEEE Transactions on Software Engineering*, 31(1):21–37, January 2005.

[Joerg, 2005] Evermann Joerg. The association construct in conceptual modelling-an analysis using the bunge ontological model. In *In Proc. of The 17th Conference on Advanced Information Systems Engineering (CAiSE 2005) LNCS 3520*, pages 33–47, June 2005.

[Michael and Green, 2002] Rosemann Michael and Peter Green. Developing a meta model for the bunge-wand-weber ontological constructs. *Information Systems*, 27:75–91, 2002.

[OMG, ] OMG. *Unified Modeling Language*. Object Management Group.

[Sowa, 2000] J. F. Sowa. *Knowledge Representation: Logical, Philosophical, and Computational Foundations*. Brooks Cole Publishing Co., 2000.

[Uschold *et al.*, 1998] Mike Uschold, M. King, S. Moralee, and Y. Zorgios. Enterprise ontology. *The Knowledge Engineering Review*, 13(1):31–89, 1998.

[Yair and Weber, 1990] Wand Yair and Ron Weber. An ontological model of an information system. *IEEE Transactions on Software Engineering*, 16(11):1282–1292, November 1990.

[Yair and Weber, 1993] Wand Yair and Ron Weber. On the ontological expressiveness of information system analysis and design grammars. *Journal of Information Systems*, (3):217–237, 1993.

[Yair and Weber, 1995] Wand Yair and Ron Weber. On the deep structure of information systems. *Information System Journal*, (5):203–223, 1995.

[Yair and Weber, 1999] Wand Yair and Ron Weber. An ontological analysis of the relationship construct in conceptual modeling. *ACM Transactions on Database Systems*, 24(4):494–528, December 1999.